\documentclass{article}

\usepackage{arxiv}

\usepackage[utf8]{inputenc} 
\usepackage[T1]{fontenc}    
\usepackage{hyperref}       
\usepackage{url}            
\usepackage{booktabs}       
\usepackage{amsfonts}       
\usepackage{nicefrac}       
\usepackage{microtype}      
\usepackage{lipsum}		
\usepackage{graphicx}
\usepackage{natbib}
\usepackage{doi}
\usepackage{tabularx} 
\usepackage{amsmath} 
\usepackage{array}
\usepackage{multirow}

\usepackage{dcolumn}
\usepackage{caption}
\usepackage{amsmath}
\usepackage{amssymb}

\title{An Empirical Analysis of Tiff's Impact on American Business Formation}

\author{ \href{https://raymonmin.github.io}{
        \hspace{1mm}Ruiming MIN}\\
	Department of Mathematics\\
        The Hong Kong University of Science and Technology\\
	\texttt{rmin@connect.ust.hk} \\
}

\hypersetup{
pdftitle={An Empirical Analysis of Tiff's Impact on American Business Formation},
pdfauthor={Ruiming MIN},
pdfkeywords={Trump tariffs, US-China trade war, business formation, economic revitalization, protectionism, retaliatory tariffs, manufacturing jobs, trade policy, economic impact analysis, MAGA policy, business applications, county-level data, linear regression analysis, Peterson Institute, Census Bureau data},
}

\begin{document}
\maketitle

\begin{abstract}
This study examines whether the tariff policies, particularly the US-China trade war initiated in 2018, delivered on promises to revitalize American manufacturing and create jobs. Using county-level business application data from 2018-2025, we analyze the relationship between tariff implementation and new business formation through linear regression analysis. Our findings reveal a statistically significant positive association between US tariffs on China and American business applications. However, when Chinese retaliatory tariffs are included in the analysis, their negative coefficient substantially exceeds the positive US tariff effect, suggesting that retaliatory measures largely offset the benefits of protectionist policies. Control variables including inflation rate, federal funds rate, and government spending show significant positive effects on business formation. These results indicate that while protectionist trade policies may stimulate domestic business formation, their effectiveness is significantly diminished by retaliatory responses from trading partners. The study provides evidence that unilateral tariff measures without diplomatic coordination produce limited net benefits, confirming that trade wars create scenarios where potential gains are neutralized by counteractions.

\end{abstract}


\section{Introduction}

Since 2016, Donald Trump has employed the slogan "Make America Great Again" (MAGA) to rally public support and win his election. One of his most impressive policies promised to his supporters was his tariff policy, particularly his trade war with China since 2018. In his speech, he asserted that his tariffs would lead to a return of American industry and manufacturing jobs. "Jobs and factories will come roaring back into our country," he said while announcing his tariffs \citep{abcnews2025}. 

This raises a fundamental question: Did his policy actually deliver the promised outcomes?
In previous research, \citet{fajgelbaum2020return} analyzed the short-term impact of 2018 U.S. tariff policies, finding that tariffs caused significant declines in imports and exports, with complete pass-through of tariffs to U.S. consumers.
Similarly, research in the Federal Reserve Board finds that its positive effect from import protection is offset by larger negative effects from rising input costs and retaliatory \citep{amiti2019impact}. 
Moreover, there is plenty of research leading to similar result indicating the effect on various aspects of both the U.S. and China economy, such as \citet{10.1257/pandp.20201018}, \cite{amiti2019impact}, \citet{NBERw26353}, \cite{flaaen2019disentangling}, and \citet{CUI2021109846}.

However, by analyzing the relationship between his tariff policies and business application rates in the United States, we find evidence that contradicts the wholly negative assessments using county-level data from 2018-2025 from \citet{CensusBFS2025}, \citet{fred2025}, and \citet{PIIE2019}. Our results suggest that these tariffs may have actually stimulated new business formation, partially fulfilling Trump's promise of economic revitalization, and provide another view of Trump's tariffs for further exploration.

\section{Data and Methods}

In our work, we only consider the data from 2018 to 2025, as the trade war between the U.S. and China began in 2018. This range will contain the full evolution of the trade war, from the initial Section 301 tariffs implemented in July 2018 through subsequent escalations, allowing us to analyze both immediate and medium-term economic impacts.

The most recent data is from April 2025. This 88-month panel provides sufficient variation to examine how tariff shocks propagate through various economic channels over time. The extended timeframe also enables us to control for other macroeconomic factors that might influence the relationship between trade policy and economic outcomes.

We consider the business application for all NAICS from \citet{CensusBFS2025} as the relevant variable. Meanwhile, the U.S. Tariffs on China from \citet{PIIE2019} will be the independent variable we are most interested in. Since the tariff may change multiple times in one month, we will use a weighted sum of the tariff by the number of continuing days every month. 

In terms of the control variables we used, we chose the unemployment rate to exclude those necessity entrepreneurs due to the economic shock. We add the inflation rate and the Federal Funds Rate to reflect the effect of the market environment as well. As we also want to isolate the effect of the government force in this case, we use government spending to control it. All of the data above are from \citet{fred2025}. The data on government spending is only provided in quarterly reports. Thus, we simply divided each by $3$ to fit the monthly data.

Moreover, there is one special control variable we want to mention, which is China's Tariffs on the U.S. from \citet{PIIE2019}, as China was the largest trading partner before the trade war. We can also find some fun facts by adding this control variable, and we will discuss it in the following sections.

In order to get the most intuitive result of the tariff, we introduce the simplest linear model to fit the regression analysis through time:
\begin{align*}
    & \#{Business\_Application} \\
    = & \beta_0 + \gamma Z_{US\_Tariffs\_on\_CN} + \beta_1 X_{unemploy} + \beta_2 X_{inflation} + \beta_3 X_{DFF} + \beta_4 X_{Government} \\
    & ( + \beta_5 X_{CN\_Tariffs\_on\_US})
\end{align*} 
where $\gamma$ refers to the effect of Trump's tariffs.

One major concern about our data and method may be the range of our data, which includes Biden's term. However, although Biden didn't clamor for increasing tariffs or threaten other countries with tariffs, he didn't decrease Trump's tariffs either. Thus, we consider including his term to be reasonable and necessary, which is used in other articles like \citet{Bown2022China} and \citet{Bown2022Four}.

Moreover, since we only want to obtain a rough qualitative analysis, the linear model is enough to provide the relationship between variables. But the value of our model's coefficient is not very important and can't be used in the prediction process. The sign and significance level would be the most important term.

\section{Result and Discussion}

Columns 1-3 and Columns 4-6 in Table \ref{tab:result} report the regression results without and with Chinese tariff variables, respectively. Overall, we find significant relationships between US tariff increases during the trade war and our study variable. Specifically, without considering Chinese tariffs, a one percentage point increase in US tariffs on China is associated with a 1819 unit increase in the dependent variable (Column 1), which is significant at the 1\% level. After including Chinese tariffs on the U.S., the coefficient of US tariffs increases to 5142 (Column 4), though its significance level drops to 10\%. This decrease in significance level is reasonable, as the correlation between these two variables is quite high. 

The positive coefficient of  $Z_{US\_Tariffs\_on\_CN}$ provides us with an unexpected result compared with the result by \citet{amiti2019impact}. Our regression indicates that Trump's tariffs do have a positive effect on the return of American firms in terms of the number of new firms. However, if we consider the comparison between the two models, the result can be explained as the counterattack by China and other countries offsetting the positive impact of Trump’s tariffs, since the absolute value of  $Z_{US\_Tariffs\_on\_CN}$'s coefficient is less than the absolute value of $X_{CN\_Tariffs\_on\_US}$'s coefficient. Then, our result would align with the other article's result, like \citet{amiti2019impact}.

\begin{table}[!htb]
\centering
\begin{tabular}{lcccccc}
\toprule
& \multicolumn{3}{c}{Regression w/o Chinese Tariff} & \multicolumn{3}{c}{Regression w Chinese Tariff} \\
\cmidrule(lr){2-4} \cmidrule(lr){5-7}
Variable & Coef. & Std. Err. & p-value & Coef. & Std. Err. & p-value \\
\midrule
Intercept & 128400$^{***}$ & (40900) & 0.002 & 171200$^{***}$ & (55680) & 0.003 \\
US Tariffs on CN & 1819$^{***}$ & (681.7) & 0.009 & 5142$^{*}$ & (3019) & 0.092 \\
CN Tariffs on US & -- & -- & -- & -5296 & (4687) & 0.262 \\
Unemployment Rate & 5563 & (5547) & 0.319 & 5958 & (5549) & 0.286 \\
Inflation & 1.226e7$^{***}$ & (3.205e6) & 0.000 & 1.266e7$^{***}$ & (3.219e6) & 0.000 \\
Federal Funds Rate & 13610$^{**}$ & (5352) & 0.013 & 13470$^{**}$ & (5344) & 0.014 \\
Gov. Spending & 64.93$^{***}$ & (19.73) & 0.001 & 65.05$^{***}$ & (19.70) & 0.001 \\
\midrule
R-squared & \multicolumn{3}{c}{0.429} & \multicolumn{3}{c}{0.438} \\
Adj. R-squared & \multicolumn{3}{c}{0.394} & \multicolumn{3}{c}{0.396} \\
F-statistic & \multicolumn{3}{c}{12.33$^{***}$} & \multicolumn{3}{c}{10.52$^{***}$} \\
Observations & \multicolumn{3}{c}{88} & \multicolumn{3}{c}{88} \\
\bottomrule
\multicolumn{7}{l}{\footnotesize Note: $^{*}$ p $<$ 0.1, $^{**}$ p $<$ 0.05, $^{***}$ p $<$ 0.01. Standard errors in parentheses.} \\
\end{tabular}
\caption{Regression Analysis Results}
\label{tab:result}
\end{table}

Among other control variables, inflation rate, federal funds rate, and government spending all show significant positive effects, with inflation rate and government spending being significant at the 1\% level and federal funds rate at the 5\% level. These results align with macroeconomic theory, which shows that the rise of inflation leads to an increase in investment. Similarly, the higher government spending directly stimulates the government's power in supporting local firms. While the unemployment rate shows a positive coefficient, it is not statistically significant, which weakly indicates the necessity entrepreneurship. The F-statistics significant at the 1\% level, indicating good model fit.

\section{Conclusion}
This paper examines whether Donald Trump's tariff policies, particularly the US-China trade war initiated in 2018, delivered on his promise to revitalize American manufacturing and bring back jobs. Through an analysis of US business application data from 2018-2025, the research offers a nuanced perspective on tariff effectiveness. The study reveals that contrary to wholly negative assessments in previous research, Trump's tariffs demonstrated a statistically significant positive association with new business formation in the United States. However, when Chinese retaliatory tariffs were included in the analysis, the picture became more complex. The absolute value of the Chinese tariff coefficient exceeded that of US tariffs, suggesting that the retaliatory measures may have substantially offset the positive impacts of Trump's policy. 

These findings indicate that protectionist trade policies may stimulate domestic business formation, but their effectiveness is significantly diminished when trading partners implement retaliatory measures. This confirms the economic theory that trade wars create lose-lose scenarios where potential gains from protective tariffs are neutralized by counteractions. For policymakers considering tariffs as a tool for economic revitalization, this research suggests that unilateral protectionist measures without diplomatic coordination are likely to produce limited net benefits due to retaliatory responses. 

The study focused primarily on business application rates as a metric for economic revitalization, which captures only one dimension of Trump's promises. Future research should examine employment figures, wage growth, and manufacturing output to provide a more comprehensive assessment of tariff effectiveness. Additionally, the linear model employed offers a simplified view of complex economic relationships. More sophisticated econometric approaches, including structural models that can better account for industry-specific impacts and regional variations, would enhance our understanding of tariff effects. As the international trade landscape continues to evolve, longitudinal studies tracking the long-term impacts of these tariff policies will be essential for fully evaluating their economic legacy.

\bibliography{ref}

\begin{thebibliography}{}

\bibitem[\protect\citeauthoryear{{ABC News}}{{ABC News}}{2025}]{abcnews2025}
{ABC News} (2025, April).
\newblock Trump tariffs won't entice companies to expand us manufacturing,
  economic experts warn.
\newblock
  \url{https://abcnews.go.com/Politics/trump-tariffs-entice-companies-expand-us-manufacturing-economic/story?id=120635951}
  (accessed May 18, 2025).

\bibitem[\protect\citeauthoryear{Amiti, Redding, and Weinstein}{Amiti
  et~al.}{2019}]{amiti2019impact}
Amiti, M., S.~J. Redding, and D.~Weinstein (2019).
\newblock The impact of the 2018 tariffs on prices and welfare.
\newblock {\em Journal of Economic Perspectives\/}~{\em 33\/}(4), 187--210.

\bibitem[\protect\citeauthoryear{Amiti, Redding, and Weinstein}{Amiti
  et~al.}{2020}]{10.1257/pandp.20201018}
Amiti, M., S.~J. Redding, and D.~E. Weinstein (2020, May).
\newblock Who's paying for the us tariffs? a longer-term perspective.
\newblock {\em AEA Papers and Proceedings\/}~{\em 110}, 541–46.

\bibitem[\protect\citeauthoryear{Bown}{Bown}{2022a}]{Bown2022China}
Bown, C.~P. (2022a, July).
\newblock China bought none of the extra \$200 billion of {US} exports in
  {Trump's} trade deal.
\newblock {\em Peterson Institute for International Economics\/}.
\newblock Last updated: December 3, 2024.

\bibitem[\protect\citeauthoryear{Bown}{Bown}{2022b}]{Bown2022Four}
Bown, C.~P. (2022b, October).
\newblock Four years into the trade war, are the {US} and {China} decoupling?
\newblock {\em Peterson Institute for International Economics\/}.
\newblock Last updated: November 7, 2024.

\bibitem[\protect\citeauthoryear{Cui and Li}{Cui and Li}{2021}]{CUI2021109846}
Cui, C. and L.~S.-Z. Li (2021).
\newblock The effect of the us–china trade war on chinese new firm entry.
\newblock {\em Economics Letters\/}~{\em 203}, 109846.

\bibitem[\protect\citeauthoryear{Fajgelbaum, Goldberg, Kennedy, and
  Khandelwal}{Fajgelbaum et~al.}{2020}]{fajgelbaum2020return}
Fajgelbaum, P.~D., P.~K. Goldberg, P.~J. Kennedy, and A.~K. Khandelwal (2020).
\newblock The return to protectionism.
\newblock {\em The Quarterly Journal of Economics\/}~{\em 135\/}(1), 1--55.

\bibitem[\protect\citeauthoryear{{Federal Reserve Bank of St. Louis}}{{Federal
  Reserve Bank of St. Louis}}{2025}]{fred2025}
{Federal Reserve Bank of St. Louis} (2025).
\newblock Federal reserve economic data.
\newblock \url{https://fred.stlouisfed.org/} (accessed May 18, 2025).

\bibitem[\protect\citeauthoryear{Flaaen and Pierce}{Flaaen and
  Pierce}{2019}]{flaaen2019disentangling}
Flaaen, A. and J.~R. Pierce (2019, December).
\newblock Disentangling the effects of the 2018-2019 tariffs on a globally
  connected u.s. manufacturing sector.
\newblock Finance and Economics Discussion Series 2019-86, Board of Governors
  of the Federal Reserve System.

\bibitem[\protect\citeauthoryear{{Peterson Institute for International
  Economics}}{{Peterson Institute for International
  Economics}}{2019}]{PIIE2019}
{Peterson Institute for International Economics} (2019, August).
\newblock Us-china trade war tariffs: An up-to-date chart.
\newblock
  \url{https://www.piie.com/research/piie-charts/2019/us-china-trade-war-tariffs-date-chart}
  (accessed May 18, 2025).

\bibitem[\protect\citeauthoryear{{U.S. Census Bureau}}{{U.S. Census
  Bureau}}{2025}]{CensusBFS2025}
{U.S. Census Bureau} (2025).
\newblock Business formation statistics.
\newblock
  \url{https://www.census.gov/econ/currentdata/?programCode=BFS&startYear=2018&endYear=2025&categories[]=TOTAL&dataType=BA_BA&geoLevel=US&adjusted=0¬Adjusted=1&errorData=0#table-results}
  (accessed May 18, 2025).

\bibitem[\protect\citeauthoryear{Waugh}{Waugh}{2019}]{NBERw26353}
Waugh, M.~E. (2019, October).
\newblock The consumption response to trade shocks: Evidence from the us-china
  trade war.
\newblock Working Paper 26353, National Bureau of Economic Research.

\end{thebibliography}

\end{document}